\documentclass[10pt,prd,twocolumn,preprint,nofootinbib]{revtex4}

\usepackage{epsfig}
\usepackage{amsmath,amsfonts,amssymb}
\usepackage{t1enc}
\usepackage{verbatim}
\usepackage{float}
\usepackage{morefloats}

\providecommand{\openone}{\leavevmode\hbox{\small1\kern-3.8pt\normalsize1}}

\usepackage{booktabs}
\usepackage[Q=yes,pverb-linebreak=no]{examplep}

\newcommand{\Vl}{V_L}
\newcommand{\Vr}{V_R}
\newcommand{\gl}{g_L}
\newcommand{\gr}{g_R}

\newcommand{\fz}{F_0}

\newcommand{\fr}{F_R}

\newcommand{\be}{\begin{equation}}
\newcommand{\ee}{\end{equation}}

\begin{document}

\title{\boldmath Global Constraints on Top Quark Anomalous Couplings \unboldmath}

\author{
Fr\'ed\'eric~D\'eliot$^{1}$,
Ricardo~Faria$^{2}$,
Miguel C. N. Fiolhais$^{3,4}$,
Pedro~Lagarelhos$^{2}$,
Ant\'onio Onofre$^{2}$,
Christopher~M.~Pease$^{5}$,
Ana Vasconcelos$^{2}$
\\[3mm]
{\footnotesize {\it 
$^1$ Institut de Recherche sur les lois Fondamentale de l'Univers, Service de Physique des Particules, CEA Saclay - Bat 141, F-91191 Gif-sur-Yvette Cedex, France \\
$^2$ LIP, Departamento de F\'{\i}sica, Universidade do Minho, 4710-057 Braga, Portugal \\
$^3$ Science Department, Borough of Manhattan Community College, City University of New York, \\ 199 Chambers St, New York, NY 10007, USA \\
$^4$ LIP, Departamento de F\'{\i}sica, Universidade de Coimbra, 3004-516 Coimbra, Portugal \\
$^5$The Graduate School and University Center, The City University of New York, 365 Fifth Avenue, New York, NY 10016  USA \\
}}
}


\begin{abstract}
The latest results on top quark physics, namely single top quark production cross sections, $W$-boson helicity and asymmetry measurements are used to probe the Lorentz structure of the $Wtb$ vertex. The increase of sensitivity to new anomalous physics contributions to the top quark sector of the Standard Model is quantified by combining the relevant results from Tevatron and the Large Hadron Collider. The results show that combining an increasing set of available precision measurements in the search for new physics phenomena beyond the Standard Model leads to significant sensitivity improvements, especially when compared with the current expectation for the High Luminosity run at the LHC.
\end{abstract}

\maketitle

\section{Introduction}
\label{sec:intro}

Precision measurements of observables sensitive to the $Wtb$ vertex structure provide an important test to the Standard Model (SM). Recent results at the Large Hadron Collider (LHC) show an increasingly good agreement between the experimental measurements and the high precision theoretical predictions within the SM. This implies that the search for new anomalous physics contributions to this vertex will naturally require new innovative strategies. As such, the introduction of new observables and the implementation of high precision measurements in the forthcoming runs at the LHC will play an important role to still increase the sensitivity to possible new physics beyond the SM (BSM). 

New physics effects above the electroweak symmetry breaking scale on the $Wtb$ vertex can be probed in the framework of an effective field theory approach~\cite{Buchmuller:1985jz}. In this approach, the most general $Wtb$ vertex lagrangian including dimension-six gauge-invariant effective operators can be expressed as~\cite{AguilarSaavedra:2008zc,AguilarSaavedra:2009mx}:
\begin{eqnarray}
\mathcal{L}_{Wtb} & = & - \frac{g}{\sqrt 2} \bar b \, \gamma^{\mu} \left( \Vl
P_L + \Vr P_R
\right) t\; W_\mu^- \nonumber \\
& - &  \frac{g}{\sqrt 2} \bar b \, \frac{i \sigma^{\mu \nu} q_\nu}{M_W}
\left( \gl P_L + \gr P_R \right) t\; W_\mu^- + \mathrm{h.c.}  
\label{ec:lagr}
\end{eqnarray}
In the SM at tree level, the left-hand vectorial coupling $\Vl$ is determined by the CKM matrix element $V_{tb} \simeq 1$, while the anomalous couplings, $\Vr$, $\gl$, $\gr$ are equal to zero. These couplings are dimensionless, complex, and may receive contributions at higher orders in the SM and in the presence of BSM physics.

Anomalous contributions to the $Wtb$ vertex have been systematically probed in the scientific literature~\cite{Aaboud:2016hsq,Birman:2016jhg,Castro:2016jjv,Bernardo:2014vha,Aaboud:2017yqf}. The $W$ boson helicity fractions in top quark decays, and their corresponding angular asymmetries, are particularly sensitive to anomalous couplings. These observables have been measured by Tevatron and LHC experiments in different channels of top quark pair and single top quark productions, and used to establish limits on the aforementioned anomalous couplings. Single top quark production cross-section measurements have also been combined with $W$ boson helicity fractions in current and prospective scenarios, and proved to be extremely efficient in significantly improving allowed regions for anomalous contributions~\cite{Boos:1999dd,Najafabadi:2008pb,AguilarSaavedra:2008gt,Chen:2005vr,Zhang:2010dr,AguilarSaavedra:2010nx}. 

Furthermore, useful spin directions perpendicular to the $W$ boson momentum ($\vec{p}_W$) were recently introduced in the production of polarized top quarks on the $t$-channel single top quark electroweak process at the LHC ~\cite{AguilarSaavedra:2010nx, Aguilar-Saavedra:2015yza}. These are particularly relevant as they allow the definition of new angular asymmetries, sensitive to the complex phases of the $Wtb$ vertex anomalous couplings. In particular, the normal direction,
\begin{equation}
\vec{N}=\vec{s}_t\times \vec{p}_W \, , 
\end{equation}
perpendicular to the plane defined by $\vec{p}_W$ and the top spin direction ($\vec{s}_t$), and the transverse direction,
\begin{equation}
\vec{T}=\vec{p}_W \times \vec{N} \, , 
\end{equation}
were introduced, leading to the angles $\theta^{N}_{\ell}$ ($\theta^{T}_{\ell}$) between the charged lepton momentum in the $W$ center-of-mass system and the normal (transverse) direction in the top quark rest frame, which were used to define the $A_{FB}^{N}$ ($A_{FB}^{T}$) asymmetries. In addition, by considering the angle between the lepton momentum in the top quark rest frame and the top quark spin direction ($\theta_{\ell}$), the forward-backward asymmetry  $A_{FB}^{\ell}$ was also introduced.  These asymmetries have been measured by the ATLAS collaboration at the LHC~\cite{Aaboud:2017aqp}, in good agreement with the SM prediction.

In this paper, the limits on anomalous contributions to the $Wtb$ vertex are calculated using the latest and most precise top quark measurements at the LHC and Tevatron. The latest results on the measurements of the $W$-boson helicity fractions, angular asymmetries and spin observables are combined with single top quark production cross section measurements in order to establish allowed regions at 95\%~confidence level (CL). These results are presented in different scenarios in order to determine the most efficient way to increase the sensitivity to new physics phenomena on the $Wtb$ vertex.

\section{$W$ Boson Helicity Fractions}
\label{sec:measurements}


The most precise longitudinal and left-handed $W$ boson helicity fraction measurements obtained at the LHC were considered in the global fit of the anomalous couplings~\cite{Aaboud:2016hsq},
\begin{eqnarray}
F_0 & = & 0.709 \pm 0.019 \, , \\ 
F_L & = & 0.299 \pm 0.015 \,.
\end{eqnarray}
The two helicity fractions were measured to be anti-correlated with an overall correlation coefficient of \mbox{$\rho=-0.55$}. The Tevatron results were also taken into account in the fit, 
\begin{eqnarray}
\fz & = & \phantom{-}0.722 \pm 0.081 \, , \\ 
\fr & = & -0.033\pm0.046  \, ,
\end{eqnarray}
with a correlation coefficient of \mbox{$\rho = -0.86$}~\cite{Aaltonen:2012rz}. 

The relative differences between the helicity fraction measurements and the most accurate SM predictions at next-to-next-to-leading order (NNLO) in QCD~\cite{Czarnecki:2010gb}, 
\begin{eqnarray}
F_0 & = & \phantom{7}0.687 \pm 0.005 \, , \\
F_L & = & \phantom{7}0.311 \pm 0.005 \, , \\ 
F_R & = & 0.0017 \pm 0.0001 \, , 
\end{eqnarray}
normalized to the total experimental uncertainty (statistical and systematical), are shown in Figure~\ref{fig:observables}.

\section{Single Top Quark Cross Sections}
\label{sec:measurements2}

\begin{figure*}
\begin{center}
\includegraphics[width=12.2cm]{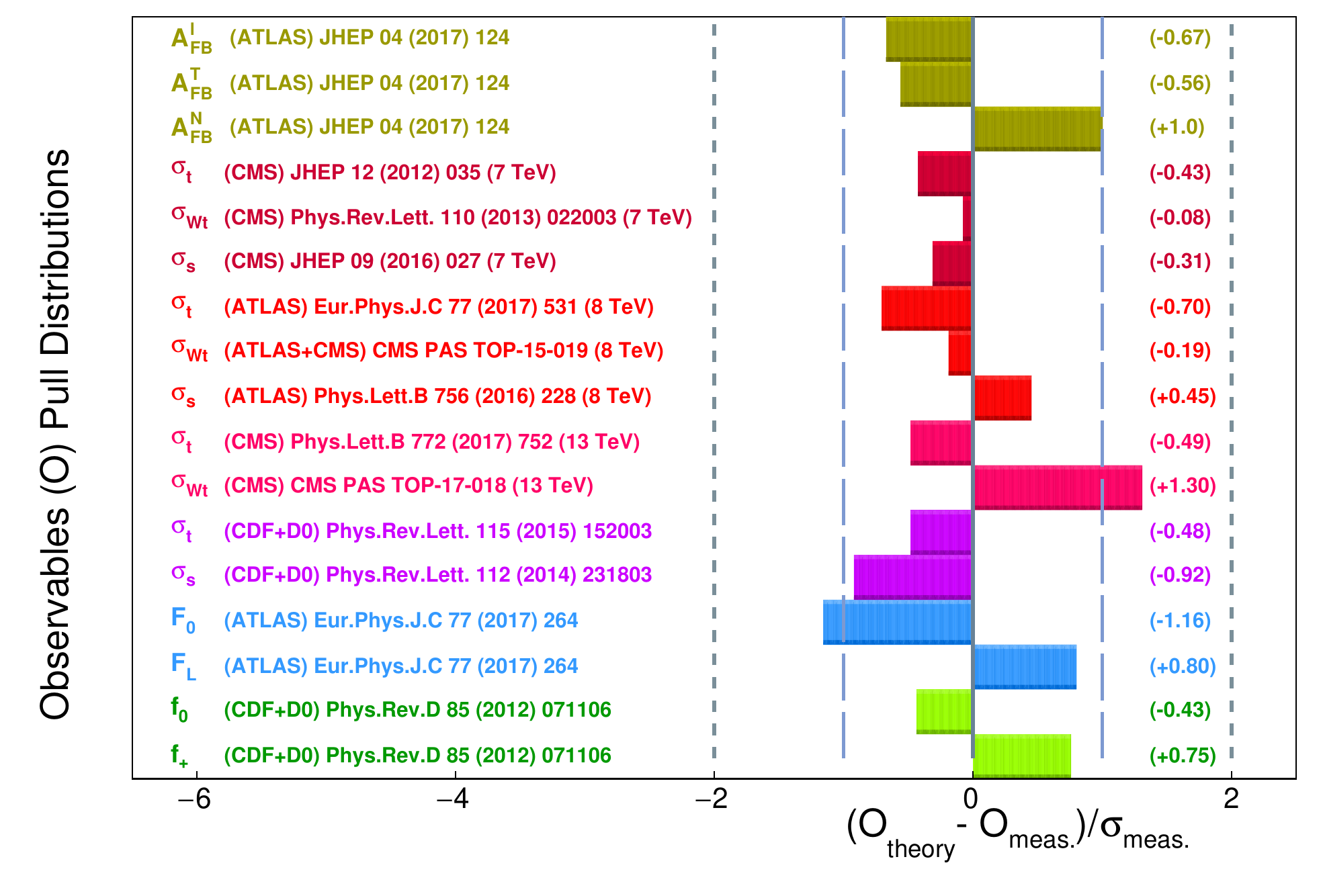}
\caption{Relative difference between observable measurements and their most accurate SM predictions.}
\label{fig:observables}
\end{center}
\end{figure*}

The experimental measurements on single top quark production at the Tevatron and LHC, through the electroweak $t$-, $Wt$- and $s$-channels were also considered, at different center-of-mass energies. For instance, at the LHC, the most precise measurements on the $t$-channel and $Wt$ associated productions at $\sqrt{s}=$13~TeV, are 
\begin{eqnarray}
\sigma_t&=&232\pm 31~\textrm{pb} \, , \\
\sigma_{Wt}&=&63.1\pm1.8 \, \textrm{(stat)}\pm6.0 \, \textrm{(syst)} \,\nonumber \\
           & & \pm  2.1 \, \textrm{(lumi)}~\textrm{pb} \, ,
\end{eqnarray}
respectively~\cite{Sirunyan:2016cdg,Khachatryan:2017-018}. The $t$-channel cross section measurement is compared with the calculation at next-to-leading order (NLO) from HATHOR~\cite{Kant:2014oha} in Figure~\ref{fig:observables}. The NLO result, although less precise, is within few percent the NNLO~\cite{Brucherseifer:2014ama} calculation. The $Wt$ production cross section measurement is compared with the theoretical prediction computed at approximate next-to-next-to-leading-order~\cite{Kidonakis:2015nna}. Previous approximate NNLO results have also been obtained for all single top production channels by employing soft-gluon resummation techniques~\cite{Kidonakis:2011wy,Kidonakis:2010ux,Kidonakis:2010tc,Kidonakis:2012rm}.

The single top quark production cross sections measured at the LHC at $\sqrt{s}=$8~TeV were also used in this study~\cite{Aaboud:2017pdi,Khachatryan:2016-023,Aad:2015upn},
\begin{eqnarray}
\sigma_t & = & 89.6^{+7.1}_{-6.3}~\textrm{pb}\, , \\
\sigma_{Wt} & = & 23.1\pm3.6~\textrm{pb}\, , \\
\sigma_s & = & 4.8\pm0.8 \, \textrm{(stat)}^{+1.6}_{-1.3}\, \textrm{(syst)}~\textrm{pb} \, .
\end{eqnarray}
The SM theoretical predictions at 8~TeV used in Figure~\ref{fig:observables} 
were extracted at NLO from HATHOR~\cite{Kant:2014oha} for the $t$-channel and 
at NLO in QCD with resummation of next-to-next-to-leading soft gluon terms (NNLL)~\cite{Kidonakis:2011wy,Kidonakis:2010ux,Kidonakis:2010tc,Kidonakis:2012rm} for both the $Wt$ and $s$-channel. 

The considered single top quark measurements at a center-of-mass energy of $\sqrt{s}=$7~TeV at the LHC were 
\begin{eqnarray}
\sigma_t & = & 67.2\pm6.1~\textrm{pb} \, , \\
\sigma_{Wt} & = & 16^{+5}_{-4}~\textrm{pb} \, , \\
\sigma_s & = & 7.1\pm 8.1~\textrm{pb} \, ,
\end{eqnarray}
for the $t$-channel~\cite{Chatrchyan:2012ep}, for the $Wt$ associate production~\cite{Chatrchyan:2012zca} and for the $s$-channel~\cite{Khachatryan:2016ewo}, respectively. The cross sections relative differences shown in Figure~\ref{fig:observables}, took into account the theoretical  SM prediction at approximate NNLO for the $t$-channel~\cite{Kidonakis:2011wy}, the $Wt$ associate production~\cite{Kidonakis:2010ux} and 
$s$-channel~\cite{Kidonakis:2012rm}, at the LHC, for a center-of-mass energy of 7~TeV.

The  latest measurements of the single top quark production cross sections performed at the Tevatron at a center-of-mass energy of $\sqrt{s}=1.96$~TeV were also taken into account, for the $t$- and $s$-channels~\cite{CDF:2014uma,Aaltonen:2015cra},
\begin{eqnarray}
\sigma_t&=&2.25^{+0.29}_{-0.31}~\textrm{pb} \, , \\
\sigma_s&=&1.29^{+0.26}_{-0.24}~\textrm{pb} \, .
\end{eqnarray}
In Figure~\ref{fig:observables}, these results are compared with the SM prediction at approximate NNLO for the $t$-channel~\cite{Kidonakis:2011wy} and $s$-channel~\cite{Kidonakis:2010tc}.

\section{Asymmetries}
\label{sec:measurements3}


Several angular asymmetries have been measured at the LHC, using data collected at a center-of-mass energy of 8~TeV~\cite{Aaboud:2017aqp}. In particular, the normal and transverse polarization the forward-backward asymmetries  $A_{FB}^{N}$ and $A_{FB}^{T}$ are notably sensitive to anomalous contributions.  Both of these observables are sensitive to the real and imaginary parts of the anomalous coupling $g_R$, and $A_{FB}^{N}$ is particularly useful to constrain the imaginary component of this coupling. In fact, for small $g_R$ and assuming $V_L=1$ and $V_R=g_L=0$, the dependence of $A_{FB}^{N}$ with the imaginary component of $g_R$ can be written as,
\begin{equation}
A_{FB}^{N}=0.64\times P \times Im(g_R)
\end{equation}
where $P$ corresponds to the degree of polarization of the top quark. In addition, the forward-backward asymmetry $A_{FB}^{\ell}$ can also be used to constrain the value of the top quark polarization, $P$. 

The current most precise results for these asymmetry measurements are~\cite{Aaboud:2017aqp}: 
\begin{eqnarray}
A_{FB}^{N} & = & -0.04\pm0.04 \, , \\
A_{FB}^{T} & = & \phantom{-}0.39\pm0.09 \, , \\
A_{FB}^{\ell} & = &\phantom{-} 0.49\pm0.06 \, .
\end{eqnarray}
These values are also compared in Figure~\ref{fig:observables} with the expected SM predictions~\cite{AguilarSaavedra:2010nx, Aguilar-Saavedra:2015yza}.

\section{Results}
\label{sec:results}

The theoretical dependences of the $W$ helicity fractions ($F_0, F_L, F_R$), the single top quark production cross sections ($t$-, $Wt$- and $s$-channels) and the forward-backward asymmetries ($A_{FB}^{N}, A_{FB}^{T},A_{FB}^{\ell}$), with the anomalous couplings were used to extract new 95\%~CL limits on the anomalous couplings allowed regions with {\sc TopFit}~\cite{topfit}. The limits set by {\sc TopFit} assume a top quark, $W$-boson, and bottom quark mass of $m_t = 175$~GeV, $M_W = 80.4$~GeV, and $m_b = 4.8$~GeV, respectively. In this study, the real and imaginary components of the complex couplings were assumed to be non-vanishing and allowed to vary simultaneously.

\begin{figure*}
\begin{center}
\includegraphics[width=8.9cm]{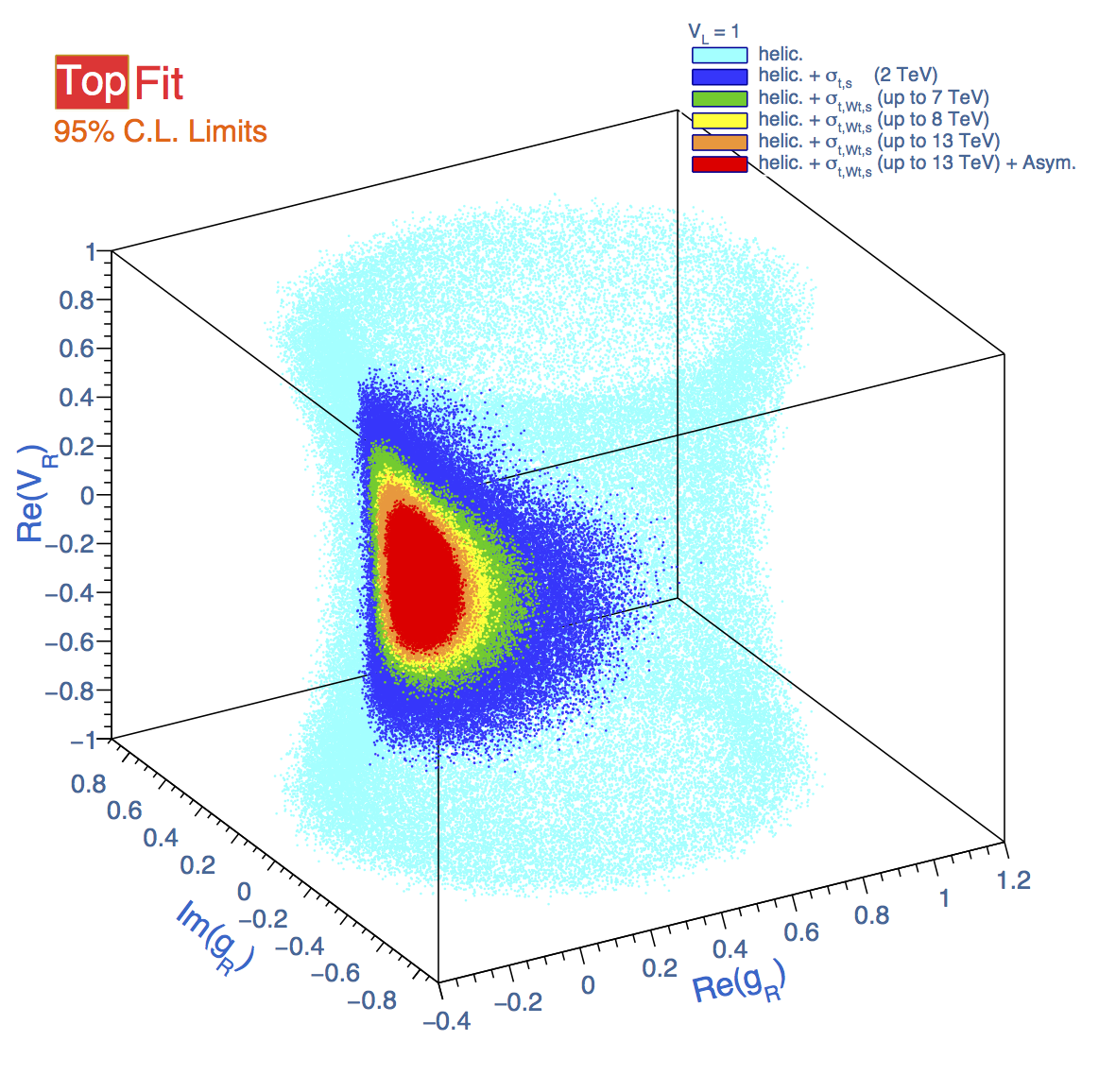}
\includegraphics[width=8.9cm]{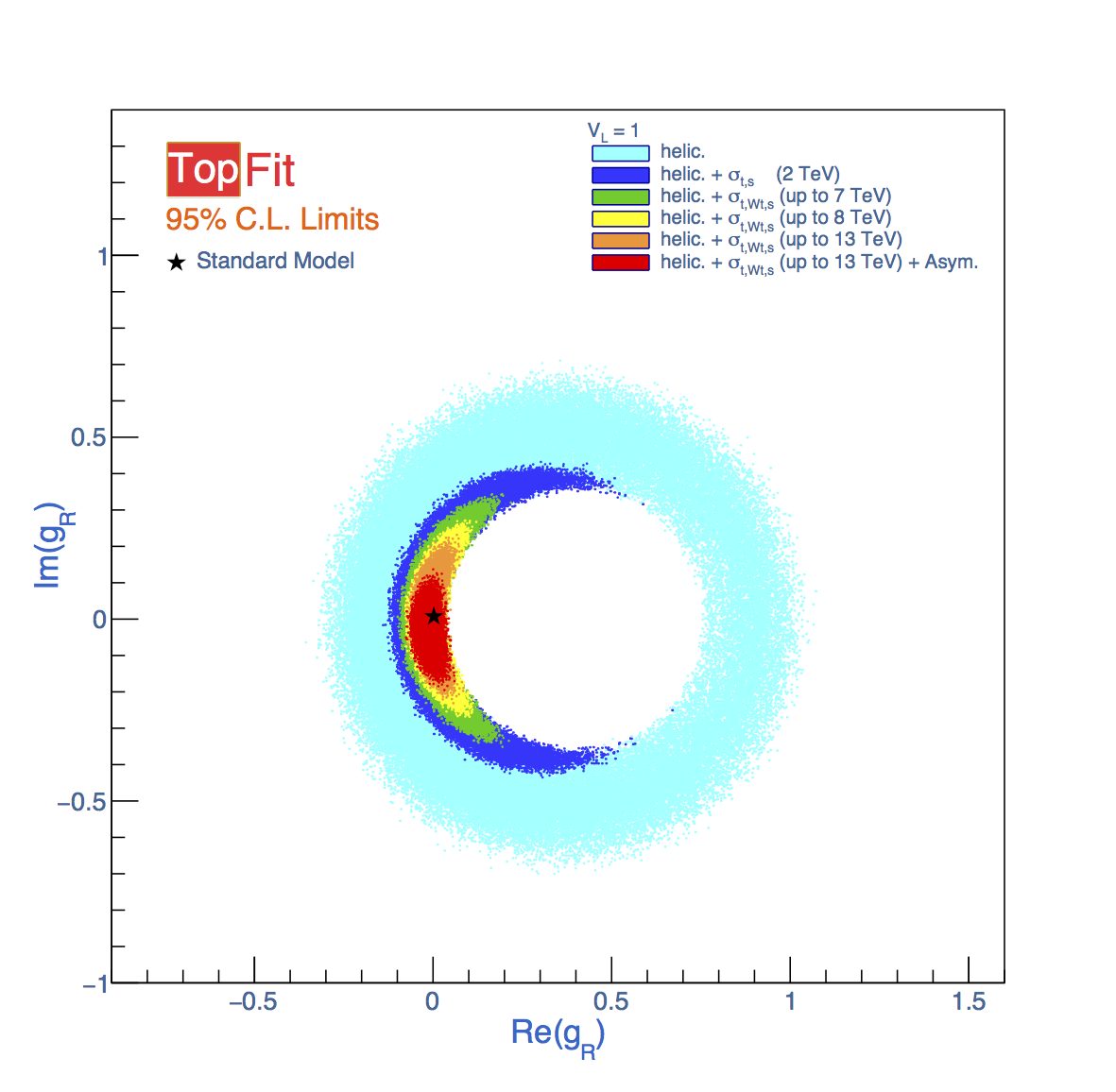}
\caption{Three-dimensional representation of the 95\%~CL limits on the allowed regions for $Re(g_R)$, $Im(g_R)$, $Re(V_R)$ (left) and the corresponding two-dimensional regions of $Re(g_R)$ as a function of  $Im(g_R)$ (right).}
\label{fig:anomalous}
\end{center}
\end{figure*}

The left plot of Figure~\ref{fig:anomalous} shows the three-dimensional representation of the 95\%~CL limits on the allowed regions for $Re(g_R)$, $Im(g_R)$, $Re(V_R)$. The plot on the right side shows the corresponding two-dimensional regions for $Re(g_R)$ and $Im(g_R)$. The different colors correspond to different sets of observables used in the global fit, in order to show the effect of progressively including additional measurements. In the light blue regions, only the $W$-boson helicity fractions are used (and assuming $|V_R|\le1$). The single top quark production cross sections measured at 1.96~TeV and up to 7, 8 and 13~TeV are also included in the dark blue, green, yellow and orange regions, respectively. It should be stressed that the latest measurement of the single top quark $Wt$ cross section (with an overall precision of about 10\%), for a center-of-mass energy of 13~TeV, at the LHC, has a strong impact on the limits. The red region in Figure~\ref{fig:anomalous} corresponds to the 95\%~CL limits on the allowed regions including the contributions of all experimental observables ($W$ boson helicity fractions, single top quark cross sections and asymmetries). The SM value is also shown.

\begin{table}[h]
\begin{center}
\begin{tabular}{|c|c|c|c|}
\hline
        $W_{hel}\oplus \sigma_{t,Wt,s}$              &      $\gr$      	&       $\gl$       		&     $\Vr$ \\ 
\hline
     Allowed Region ($Re$)   	& [-0.07 , 0.08]  &   [-0.18 , 0.20]  	& [-0.33 , 0.41] \\
     Allowed Region ($Im$)   	& [-0.23 , 0.23]  &   [-0.20 , 0.19]  	& [-0.39 , 0.36] \\
\hline
\hline
 $\oplus A^{\ell}_{FB}, A^{T}_{FB}, A^{N}_{FB}$  &      $\gr$      &       $\gl$        &     $\Vr$ \\
\hline
     Allowed Region ($Re$)   	& [-0.07 , 0.06]  &    [-0.19 , 0.19]  & [-0.27 , 0.33] \\
     Allowed Region ($Im$)   	& [-0.19 , 0.13]  &    [-0.18 , 0.19]  & [-0.30 , 0.30] \\
\hline
\end{tabular}
\caption{95\%~CL limits on the allowed regions of the real and imaginary components of the anomalous couplings. The limits are extracted from the combination of $W$-boson helicities and single top quark production  
cross section measured at the Tevatron and the LHC (top), and including the effect of the asymmetries (bottom).}
\label{tab:ReCoup}
\end{center}
\end{table}

In order to quantify the impact of the forward-backward asymmetries, two sets of numbers are shown in Table~\ref{tab:ReCoup}. The top part of the table presents the 95\%~CL limits obtained by combining all $W$ boson helicity fractions and single top quark productions cross sections from both the Tevatron and LHC. The bottom part of the table shows the same limits taking into account the forward-backward asymmetries. As expected, the inclusion of the forward-backward asymmetries significantly improve the results, particularly the imaginary component of the couplings. In particular, the improvement on the 95\%~CL limits of the allowed intervals for $Im(g_R)$ is of the order of 30\%. Both $Re(V_R)$ and $Im(V_R)$ also improved by roughly 20\%. Moreover, a clear improvement is also observed when comparing  the combination of the full set of observables discussed in this paper with the prospected limits obtained for the high luminosity run of the LHC at 14 TeV~\cite{Birman:2016jhg}. In particular, for $Im(g_R)$, the improvement on the 95\%~CL allowed regions can be as large as 30\%.

It is also worth noting that the measurement of the triple-differential angular decay distribution in single top quarks  events provide complementary results to the limits discussed in this paper~\cite{Aaboud:2017yqf}. While the allowed regions at 95\%~CL for $Re(g_R/V_L)$ are constrained to be within~\mbox{[-0.12, 0.17]}, the corresponding limits for $Im(g_R/V_L)$ are within the interval [-0.07, 0.06]. It is clear that by combining the full set of all available measurements in the top quark sector, the limits on the allowed regions of the anomalous couplings can still be further improved in the future.

\section{Conclusions}
\label{sec:conclusions}

New limits at 95\%~CL on the allowed regions of the anomalous couplings $V_R$, $g_R$ and $g_L$, were presented in this paper. The most precise measurements of the $W$ boson helicity fractions, the single top quark production cross sections and angular asymmetries measured both at Tevatron and at the LHC were used, allowing all real and imaginary components of the couplings to vary in the global fit. The results presented show an improvement of roughly 30\% for 95\%~CL allowed intervals of $Im(g_R)$, when the forward-backward asymmetries ($A_{FB}^{N}, A_{FB}^{T},A_{FB}^{\ell}$) were included in the global fit. The corresponding improvements for both $Re(V_R)$ and $Im(V_R)$ were of the order of 20\%. The real component of the anomalous coupling $g_R$ was constrained to be within [-0.07, 0.08] at 95\%~CL, while the corresponding limit for the imaginary part of this coupling was found to be within the interval [-0.19, 0.13]. These results are a good complement to the ones obtained by the ATLAS collaboration with the triple-differential angular decay distribution in single top quarks events~\cite{Aaboud:2017yqf}, opening the door to future combination of measurements at the LHC and to update of prospected limits expected at a high luminosity run of the LHC.

\section*{Acknowledgments}
This work was supported by the PSC-CUNY Award 60061-00~48. 



\end{document}